\documentclass[article]{tMPH2e}
\usepackage{color}
\usepackage[english]{babel}
\usepackage{amsmath}
\usepackage{amsfonts}
\usepackage{amssymb}
\usepackage{makeidx}
\usepackage{caption}
\newcommand{\br}{{\bf r}}
\newcommand{\ie}{\textit{i.e.\,}}
\newcommand{\brr}{{\bf r}_1}
\newcommand{\brt}{{\bf r}_2}

\newcommand{\ttitle}{{Bonding Description of the Harpoon Mechanism}$^{\dagger}$}

\begin{document}


\markboth{\ttitle}{Rodr{\'i}guez-Mayorga, Ramos-Cordoba, Salvador, Sol{\`a} and Matito}
\thanks{$^{\dagger}$ This
paper is dedicated to Andreas Savin on the occasion of his $65^{th}$ birthday.}

\title{\ttitle}

\author{Mauricio Rodr{\'i}guez-Mayorga$^{a,b}$, Eloy Ramos-Cordoba$^{a}$, 
Pedro Salvador$^b$, \\
Miquel Sol{\`a}$^b$ and Eduard Matito$^{a,c,\ast}$\thanks{$^{\ast}$ Corresponding author. Email: ematito@gmail.com}\\
$^a$ Faculty of Chemistry, University of the Basque Country UPV/EHU,
and Donostia International Physics Center (DIPC). 
P.K. 1072, 20080 Donostia, Euskadi, Spain\\
$^b$ Institut de Qu\'imica Computacional i Cat\`alisi (IQCC)
and Departament de Qu\'imica, Univ. Girona, Campus de Montilivi s/n, Girona, Spain.\\
$^c$ IKERBASQUE, Basque Foundation for Science, 48011 Bilbao, Spain}

\maketitle

\begin{abstract}
The lowest-lying states of LiH have been widely used to develop
and calibrate many different methods in quantum mechanics. In this paper
we show that the electron-transfer processes occurring in these two
states are a difficult test for chemical bonding descriptors and can be
used to assess new bonding descriptors on its ability to recognize the harpoon mechanism.
To this aim, we study the bond formation mechanism in a series of diatomic
molecules.
In all studied electron-reorganization mechanisms, the maximal
electron-transfer variation point along the bond formation path occurs when about half electron 
has been transferred from one atom to another. If the process takes places
through a harpoon mechanism, this point of the reaction path coincides with
the avoided crossing.
The electron sharing indices and one-dimensional plots of the 
electron localization function and the Laplacian of the electron density
along the molecular axis
can be used to monitor the bond formation in diatomics and provide a distinction
between the harpoon mechanism and a regular electron-reorganization process.
\end{abstract}

\section{Introduction}

The formation of molecules through the \textit{harpoon mechanism}~\cite{polanyi:95acr} 
occurs from the interaction
of two fragments, one with a low ionization potential (IP) and another that has a large electron 
affinity (EA). The reactants approach each other and, at a certain distance, an electron from the
fragment with low IP \textit{harpoons} the fragment with large EA, giving rise to a rapid
electron-transfer process that is triggered by the Coulomb attraction exerted by the two fragments.
The interplay between the energy exchange due to the harpoon process and the Coulomb attraction
determines the distance at which the electron transfer takes place.
The harpoon mechanism was proposed by Michael Polanyi to explain the unusually large cross sections
observed in the formation of alkali halides. Since the IP of the alkali metals
is similar to the EA of halogens, the Coulomb attraction needed to 
favor the electron-transfer process is usually
small for these species and, therefore, the electron transfer occurs at large separation of
the reactants. On the other hand, if the molecule under consideration is an alkali hydride the process takes
place at short distances because the EA of hydrogen is much smaller than the IP of alkalis.\newline

Since the dissociation of ionic species in gas phase 
through the inverse harpoon mechanism usually involves 
the transition from ionic to neutral fragments 
in an electron-transfer process, the molecule is expected to change its 
electron sharing index (ESI) ---commonly known as \textit{bond order}--- from a low 
(ionic) value at equilibrium to
a larger value (when the electron transferred lies between the two fragments)
that should decay as the
fragments separate from each other. 
Ponec and coworkers~\cite{ponec:05the} demonstrated that LiH and BeH show
a maximum of the ESI in the vicinity of the avoided crossing between 
two adiabatic states. Interestingly, the ESI maximum
was obtained using the quantum theory of atoms in molecules (QTAIM) definition
of an atom, but it could not be reproduced using a Mulliken atomic partition.
Some of us have demonstrated~\cite{matito:07fd,ramos-cordoba:unp} that
many atomic partitions fail to attain this condition and suggested that
the ability to reproduce this feature could be
used as a criterion to assess the goodness of atomic partitions~\cite{matito:07fd}.\newline

Thus far, the electron-transfer process has been also monitored using several
quantities, including
the dipole~\cite{weiner:87jpc} and higher-order moments~\cite{hyams:94jcp},
the topological properties of the electron density~\cite{hernandez:00jpca}, 
the electron localization
function (ELF)~\cite{krokidis:98njc,silvi:05monat} and other descriptors of the 
electronic structure~\cite{ponec:05the,matito:07fd,garcia-revilla:11jctc,pendas:07fd}.
However, none of these methods has been able to actually follow the electron
transferred as it jumps from one atom to the other. This sudden jump
is what differentiates the harpoon mechanism from a regular electron-reorganization reaction
and, therefore, its detection would provide a means to distinguish 
these mechanisms. 
Indeed, the three-dimensional pictures 
of the ELF do not reveal a separate basin for the electron until the electron
has been already completely transferred to the other fragment~\cite{silvi:05monat}.\newline

The goal of this work is analyzing the harpoon mechanism using a plethora of 
bonding descriptors to find which of these tools can be used to characterize the reaction.
To this aim, we will analyze the dissociation of three ionic compounds that
present the electron transfer at different interatomic separations and other
small molecules, whose formation does not take place through a harpooning reaction.
The analysis of these molecular systems shall shed some light on the abilities
of some bonding tools, providing some interesting test cases that can be used
in the development of new descriptors of the electronic structure of molecules.\newline

\section{Methodology}
In this section we briefly review several bonding descriptors that will be used
to study the harpoon mechanism. Let us assume a given partition of the molecular
space into atomic regions that will be labeled $A, B, C,...$. The integration of the
electron density within an atomic region gives rise to the electron population: 
\begin{equation}
N(A)=\int_A \rho(\br)d\br
\end{equation}
where $\rho(\br)$ is the electron density. Analogously, the pair density is 
needed to define pair populations, 
\begin{equation}
N(A,B)=\int_A\int_B\rho_2(\brr,\brt)d\brr d\brt
\end{equation}
which enter the expression of the electron sharing indices (ESI) or electron
delocalization indices~\cite{bader:75jacs,fradera:98jpca,matito:07fd}
\begin{equation}
\delta(A,B)=2\left[N(A)N(B)-N(A,B)\right]=-2\text{cov}\left(N(A),N(B)\right)
\end{equation}
that are related to the covariance of the atomic populations. The ESI are
a non-integer version of the classical concept of \textit{bond order}~\cite{coulson:39prsa}
that
provides a real value in accord with the number of electron pairs shared
by two fragments.\newline

The ELF requires the calculation of the Laplacian of the same-spin pair 
density functions~\cite{becke:90jcp,silvi:03jpca,matito:06jcp}
\ie,
\begin{equation}
\text{ELF(\br)}=\frac{1}{1+D(\br)^2}
\end{equation}
where
\begin{equation}
D(\br)=
\frac{
\nabla^2_{\mathbf r_2}\left(\rho_2^{\alpha\alpha}(\mathbf r, \mathbf
r_2)+\rho_2^{\beta\beta}(\mathbf r, \mathbf r_2)\right)\vert_{\mathbf
r_2=\mathbf r}}
{2c_F\rho^{8/3}(\br)}
\end{equation}
\newline
with $c_F=\frac{3}{10}\left(3\pi^2\right)^{2/3}$. The ELF measures the extent
of electron localization, giving values close to zero for regions with 
highly delocalized electrons
and large values in nuclear regions, lone pairs and bonding electron
pairs~\cite{savin:91ang}.
Similarly, negative values of the Laplacian of the electron density
also determine molecular regions with localized electrons, sometimes providing
an electronic picture similar to that of the ELF~\cite{bader:90book,matito:09ccr}.
The three-dimensional representations of the ELF and the Laplacian have been
repeatedly used in the literature to characterize all sorts of chemical
bonds~\cite{matito:09ccr}.\newline

Information theory has been also used to characterize the bonding patterns
and the electronic structure of molecular systems~\cite{nalewajski:00jpca,nalewajski:03cpl,geerlings:11pccp}.
Among them, the most well known are Shannon entropies and Fisher information
descriptors.
The Shannon entropies are descriptors of the spread of the electronic density 
(in position or momentum space) in accord with the following 
expressions~\cite{shannon:49book,jaynes:57pr}
\begin{equation}\label{eq:sr}
{S}_{\rho} = - \int \overline{\rho}({\bf r}) \ln \overline{\rho}({\bf r}) d{\bf r}
\end{equation}               
\begin{equation}\label{eq:sp}
{S}_{\pi} = - \int \pi({\bf p}) \ln \pi({\bf p}) d{\bf p}
\end{equation}               
where $\overline{\rho}({\bf r})$ is the position space electronic density normalized 
to unity and the $\pi({\bf p}) $ 
is the momentum space electronic density also normalized to unity. 
The sum of both entropies, $S_T = S_{\rho} + S_{\pi}$, ---known as total Shannon entropy--- 
has a physical bound
imposed by the 
uncertainity principle~\cite{bialynicki-birula:75cmp}. The Shannon entropy provides 
a \textit{global} measure of the electron density spread, whereas
the Fisher Information (FI) is defined as~\cite{fisher:25pcps}
\begin{equation}\label{eq:fr}
{F}_{\rho} =  \int \frac{|\nabla \overline{\rho} ({\bf r})|^2}{\overline{\rho} ({\bf r})} d{\bf r}
\end{equation}               
for the position space and 
\begin{equation}\label{eq:fp}
{F}_{\pi} =  \int \frac{|\nabla \pi ({\bf p})|^2}{\pi ({\bf p})} d{\bf p}
\end{equation}          
for the momentum space.  Although FI quantities provide a global number per
molecule, unlikely Shannon entropies, 
their value is largely affected by small local changes of the function as
has been proved by Lopez and coworkers~\cite{lopez:10thesis}. In this sense,
the FI provides a \textit{local} measure of the density sharpness.
The position space FI increases when the heterogeneity of the system leads to 
larger changes of the position space density, whereas the 
FI in the momentum space depends on large deviations of the momentum density,
\begin{equation}
 \pi ({\bf p})= \int |\widetilde{\Psi} ({\bf p_1},{\bf p_2},...,{\bf p_N}) |^2 d{\bf p_2}d{\bf p_3}...d{\bf p_N}  \,\,\,  ,
\end{equation}
and can be obtained from the Fourier transform that connects the momentum space 
wave function with the position space one
\begin{align}
\widetilde{\Psi} ({\bf p_1},{\bf p_2},...,{\bf p_N}) =& (2 \pi) ^{-3N/2} 
\int \Psi ({\bf r_1},{\bf r_2},...,{\bf r_N}) \times \\ & \exp [-i({\bf p_1}
\cdot {\bf r_1}+{\bf p_2}\cdot {\bf r_2}+...+{\bf p_N}\cdot {\bf r_N})] d{\bf r_1}d{\bf r_2}...d{\bf r_N} \nonumber
\end{align}  
These information theory indicators have been used in the past to characterize
the electronic structure of molecules and chemical 
reactions~\cite{lopez:09jctc,esquivel:09tca,geerlings:11pccp}.
Some authors~\cite{lopez:09jctc,esquivel:09tca} have linked these concepts
of information theory to particular chemical reactivity, attributing \textit{enhanced
sensitivity to changes on the position and momentum densities along the
chemical reaction paths}~\cite{lopez:09jctc}. In this work we will examine
the description of the harpoon mechanism afforded by these indicators.

\section{Computational Details}
We study the bond dissociation mechanism of ten diatomic molecules (BH, BeH, CO,
F$_2$, LiF, LiH, H$_2$, He$_2$, N$_2$ and O$_2$) in their ground state and
the first excited state of the same symmetry and spin multiplicity 
than the ground state for LiH and LiF. 
Born-Oppeheimer, full configuration interaction (FCI) calculations of H$_2$, He$_2$, 
LiH, BH and BeH have been performed
with a modified version of Knowles and Handy program~\cite{knowles:89cpc}, whereas
Gaussian~09~\cite{g09}
was used to obtain complete active space self-consistent field (CASSCF) energies.
Namely, 
CAS(6,6), CAS(10,6), CAS(8,6), CAS(6,6) and CAS(6,6) 
single-point evaluations were performed for N$_2$, F$_2$, O$_2$, LiF and CO,
respectively. 
The aug-cc-pVDZ basis was employed in all cases.
The LiF valence CASSCF calculation presents a wave function and energy discontinuity
around the avoided crossing point~\cite{sanchez:90cpl}. The latter problems are avoided if
FCI is used or if, more conveniently, a state-average
CASSCF calculation is performed~\cite{bauschlicher:88jcp,sanchez:90cpl}. 
The state-average CAS computation produces a qualitatively correct result with 
an avoided crossing at about 4$\AA$
in contrast to the FCI results that place the avoided 
crossing at 6-6.5$\AA$~\cite{bauschlicher:88jcp}.
In this paper we have performed a four-state average CAS(6,6) with equal weights,
which gives a qualitatively correct, discontinuity-free, description of the process
that is sufficient for the purposes of this work.
A six-state average CAS(6,6) calculation was also performed for CO.\newline

The quantum-theory of atoms-in-molecules (QTAIM)~\cite{bader:90book} 
calculations have been performed with AIMall package~\cite{aimall},
and the topological fuzzy Voronoi cells (TFVC) computations
have been done using APOST-3D software~\cite{apost}.
The TFVC is a three-dimensional atomic partition based on fuzzy atoms
that produces results very similar to those obtained with QTAIM at a much lower
computational cost~\cite{salvador:13jcp}.
The pair densities (2-PD) have been obtained with DMN~\cite{dmn} program
and inputed into ESI-3D~\cite{esi3d} to produce the ESI and atomic populations.
The ELF analysis has been performed with a modified version of
the ToPMoD program~\cite{topmod} using the Hartree-Fock like approximation of
the 2-PD in the computation of the ELF values.
This approximation is needed in order
to preserve the antisymmetry property of the 2-PD
that other more sophisticated approximations
do not fulfill and provides a very accurate description
of the ELF even for stretched bonds~\cite{feixas:10jctcelf}.
The calculation of information theory quantities has been
done with the RHO-OPS code developed in our group~\cite{rhoops}.

\section{Results}

\subsection{LiH}
LiH is the smallest ionic molecule that is formed through the harpoon mechanism.
The electron affinity of hydrogen is quite different from the ionization
potential of lithium and, therefore, the electron transfer in LiH occurs 
at shorter distances than in alkali halides. 
In this section we will analyze
two $^1\Sigma^+$
adiabatic states of LiH, namely, X$^1\Sigma^+$ and A$^1\Sigma^+$.
The former state dissociates into H($^2S$)+Li($^2S$) and the latter
into H($^2S$)+Li($^2P$).
These two states involve the crossing between three diabatic states: the
ionic ground state and the two lowest-lying covalent states.
The X$^1\Sigma^+$ state results from the crossing between the first ionic and covalent 
diabatic states, showing an avoided crossing with the A$^1\Sigma^+$ curve at about 3$\AA$.
The A$^1\Sigma^+$ state is characterized by two avoided
crossings: the first one has just been described and the second avoided crossing
at about 6$\AA$ results from the
interaction between the ionic and the second covalent states.
Therefore, the X$^1\Sigma^+$ ground state is characterized by an ionic bonding at the equilibrium
but its covalent character increases as the interatomic
distance stretches.
The A$^1\Sigma^+$ state is even more interesting because as the bond stretches
one expects two changes in its character: from rather covalent to ionic and back
to covalent.
The presence of two crossings
furnishes the A$^1\Sigma^+$ curve with a characteristic flat region that 
leads to huge anharmonic
effects and anomalous spectroscopic constants~\cite{yang:82acsss}.
LiH lowest-lying states have been widely studied in the
past~\cite{mulliken:36pr,stwalley:93jpcrd,weiner:87jpc}
due to the confluence of these peculiar features, which 
make them a formidable playground to test modern electronic structure methods.
The ground state electron harpooning process has been studied by means of 
chemical bonding descriptors~\cite{ponec:05the,matito:07fd,pendas:07fd}
but, to our knowledge, the excited state has never been explored
from the chemical bonding viewpoint.\newline

\begin{figure}[h]
\begin{center}
\includegraphics[angle=270, scale=0.4]{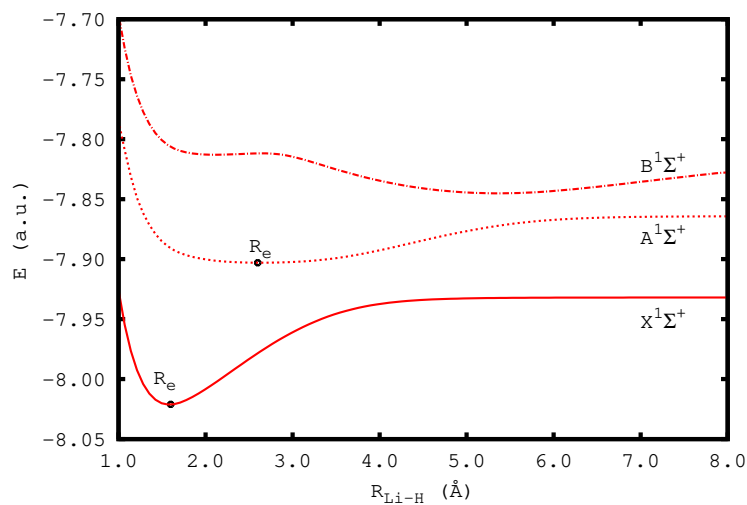}
\end{center}
\caption{Potential Energy Curve (PEC) of the LiH for the ground state (X$^1\Sigma^+$) and
the two lowest-lying $^1\Sigma^+$ excited states (A$^1\Sigma^+$ and B$^1\Sigma^+$). 
Energy in a.u.}
\label{fig:peslih}
\end{figure}

\begin{figure}[h]
\begin{center}
\includegraphics[angle=270, scale=0.4]{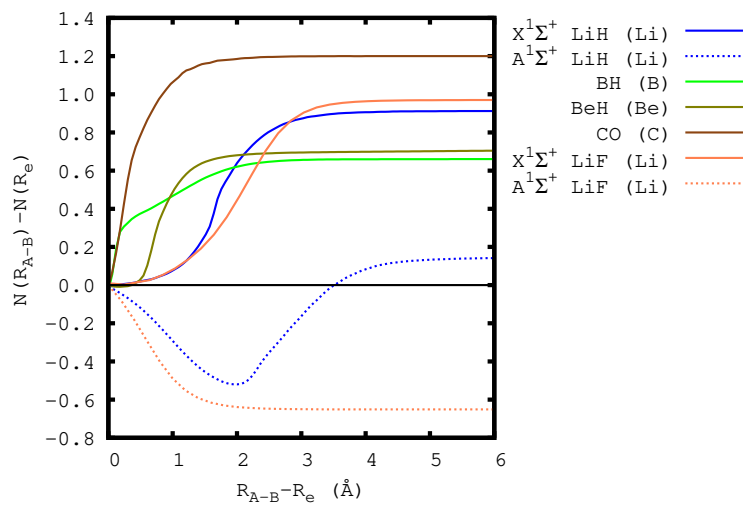}
\end{center}
\caption{Change in the atomic QTAIM population of the most electropositive
atom along the bond stretching 
in the series of studied molecules that present electron reorganization.
Population units are electrons.}
\label{fig:pops}
\end{figure}

Let us first describe the electron exchange between both atoms that occur upon stretching
the LiH bond in the two electronic states. To this aim, the most straightforward analysis
consists in checking the atomic populations along the energy curves.
Fig.~\ref{fig:pops} contains the change in the QTAIM population 
of the most electropositive atom as the molecule dissociates into two atomic fragments
(only molecules that present electron reorganization have been included).
The total density transferred from hydrogen to lithium is about 0.9 electrons
in the ground state, where the molecule passes through an avoided crossing that
changes its bonding character from ionic to covalent.
Conversely, the overall transfer
in the A$^1\Sigma^+$ state is barely 0.2 electrons. As the bond stretches the electron
passes from lithium to hydrogen until about 0.5 electrons are 
completely transferred.
This process takes places as the curve goes through
the first avoided crossing and the bond becomes more 
ionic. At about 2$\AA$ stretching, the electron transfer is reversed
and the lithium withdraws electron charge when passing by the second avoided
crossing. About 0.7 electrons are transferred in this second step and the
bond evolves from ionic to rather covalent. The covalent character 
of the molecule in the A$^1\Sigma^+$ state at geometries near
the equilibrium is not as
strong as one would expect in ground-state covalent molecules, however, the 
obtained ESI values are larger than in pure ionic molecules
and unprecedented of excited states stretched bonds (\textit{vide infra}).\newline

\begin{figure}[h]
\begin{center}
\includegraphics[angle=270, scale=0.4]{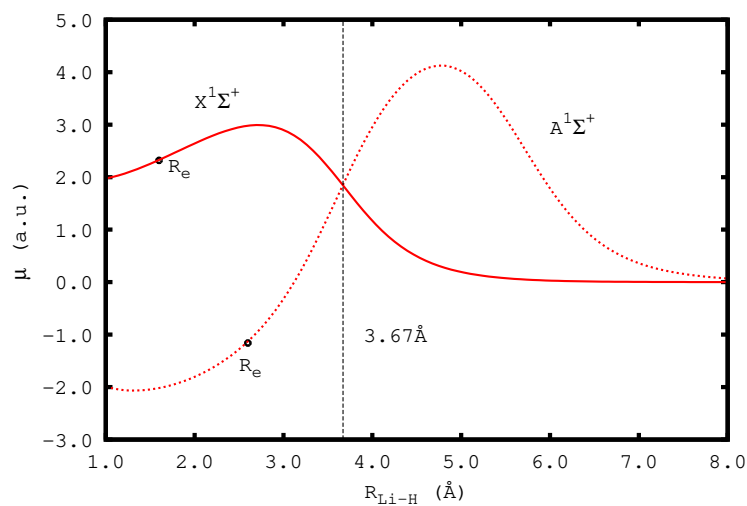}
\end{center}
\caption{Dipole moment (in a.u.) along the internuclear axis for the X$^1\Sigma^+$
and the A$^1\Sigma^+$ states of LiH as function of the interatomic distance.}
\label{fig:dipolarlih}
\end{figure}

The plot of the dipole moment 
along the internuclear axis at different interatomic separations
(Fig.~\ref{fig:dipolarlih}) can be also used to monitor the electron-transfer process.
For the ground-state at equilibrium $\mu=2.33$ a.u. and this value increases
as the bond stretches, peaking at
2.8$\AA$ and then decreasing to zero as the atoms tend to isolated neutral
entities. 
Conversely, the A$^1\Sigma^+$ state shows a lower dipole moment at the equilibrium
with opposite sign ($\mu=-1.1$ a.u.), an indication of the more covalent nature of
this bond as compared to the X$^1\Sigma^+$ state.
The dissociation of the A$^1\Sigma^+$ state of LiH involves the sign change of
the dipole moment as it passes through the first avoided crossing; reaches
a maximum value around $5\AA$ and then decreases to zero as it goes
through the second avoided crossing.\newline

\begin{figure}[hb]
\centering
\includegraphics[angle=270, scale=0.5]{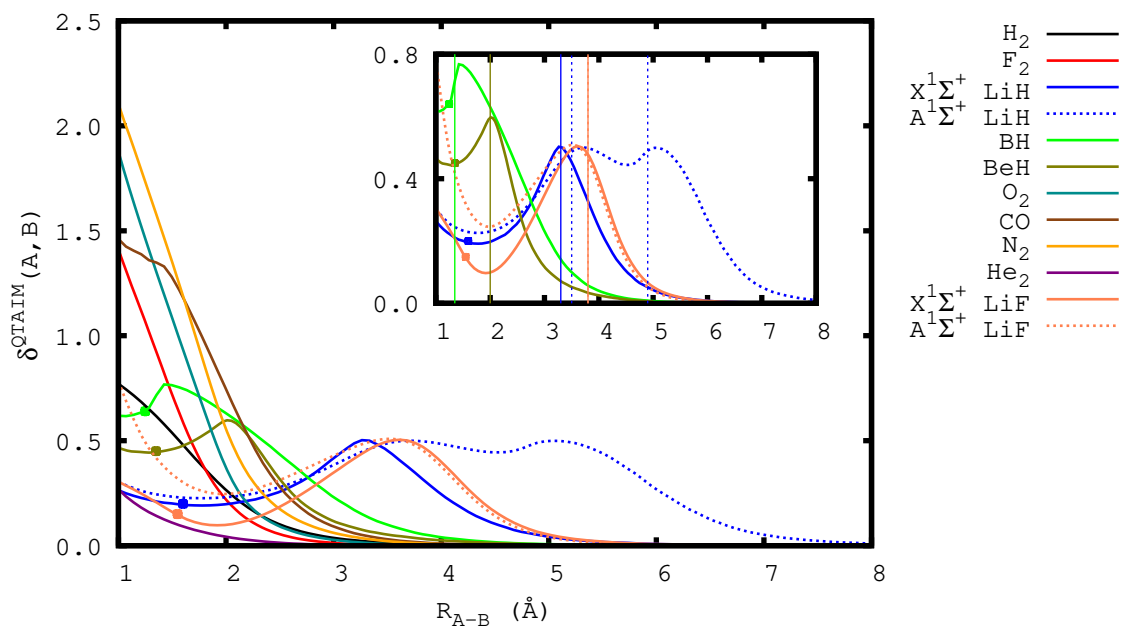}
\caption{Electron sharing indices calculated using QTAIM
partition for the series of diatomic molecules.
Solid and dotted lines are used for ground states and excited states, 
respectively. The solid points indicate the position of the equilibrium
distance and the vertical lines in the inset plot mark the position
of the maximal electron-transfer variation points (see text).
ESI units are electron pairs.}
\label{fig:esis}
\end{figure}

In Fig.~\ref{fig:esis} we find the ESIs along the dissociation
curves for a series of diatomic molecules.
The ground-state LiH ESI has a low value (0.19) at the equilibrium that 
slightly
decreases as the bond is stretched but soon enough increases as the covalent
contribution plays a more important role in the description of the X$^1\Sigma^+$
state, reaching
a maximum around $3.3\AA$ and dying off at large interatomic separations.
The ESI maximum is the point with the largest electron
sharing between the atoms, \ie, the point at which the electron is least
attached to either fragment (although, not loose enough to form a separate
entity). Indeed, the ESI peaks at the maximal transfer
variation position, \ie, the position at which the variation of electron population
with the interatomic distance 
is largest (see vertical lines in the inset plot of Fig.~\ref{fig:esis}).
Furthermore, at this point 0.5 electrons have been transferred
from Li to H when we compare their atomic populations with the ones of
the isolated neutral atoms as already reported by Pend\'as and 
coworkers~\cite{garcia-revilla:11jctc}.
The ESI can be thus used to locate the position
at which the electron remains in \textit{no one's land}, which one
identifies with the avoided crossing region. 
The analysis
of the A$^1\Sigma^+$ state reveals a similar ESI profile for interatomic
distances from 1$\AA$ to $3\AA$ than for X$^1\Sigma^+$ state. 
Despite the ESI values in this range are similar, the nature of the
bond is completely different. Let us analyze $R=1.6\AA$, which is
the equilibrium geometry for the X$^1\Sigma^+$ state. At this distance
the ground state is mostly ionic (low ESI, atomic charges close to
+1 and -1 and the dipole moment vector pointing towards hydrogen), whereas
the A$^1\Sigma^+$ state is \textit{covalent antibonding} because the charges are virtually
zero for each atom and the dipole moment vector is of the same magnitude
but opposite sign.
The LiH ESI value at equilibrium for the A$^1\Sigma^+$ state 
(${R_e^{\text{A}^1\Sigma^+}}=2.6\AA$) is larger because the molecule
becomes more covalent as it approaches the equilibrium distance
in line with the more
covalent character one expects from the excited state 
($\delta^{\text{X}^1\Sigma^+}$(Li,H)=0.19 vs.
$\delta^{\text{A}^1\Sigma^+}$(Li,H)=0.30).
For bond lengths
beyond 3$\AA$ the ESI curves of the two states deviate from each other 
and the A$^1\Sigma^+$ state presents
rather constant values (ca. 0.5) up to $5\AA$; from this point
the ESI decreases until it reaches zero beyond $8\AA$. Interestingly,
the ESI of the A$^1\Sigma^+$ state attains a second maximum around the position
of the second avoided crossing. In the formation of LiH in the
A$^1\Sigma^+$ state the two ESI maxima correspond to points at which
there is a maximal transfer variation; around the first maximum, lithium
passes electron density to hydrogen and the opposite occurs in the
vicinity of the second one. The latter is actually the point in the
molecule formation at which 0.5 electrons have been transferred from
Li to H.\newline

At this point, one should ask whether other tools to characterize the electronic
structure of molecules are capable of following the electron as it jumps from
one atom to another.
The ELF has been successfully used in the characterization of localized electrons
as those occuring in covalent bonds, lone pairs~\cite{silvi:94nat}
or even isolated electrons in ionic compounds~\cite{postils:15cc} and it has been
also used to analyze some electron-transfer processes~\cite{krokidis:98njc,silvi:05monat}.
However, the investigation of the ELF for LiH dissociation~\cite{silvi:05monat}
does not reveal a separate basin that
can be identified with the electron transferred. The process is characterized by
a progressive elongation of hydrogen's valence basin that eventually splits and
is absorbed by lithium's valence basin but, at no point,
it becomes a separate, distinctive basin.\newline

QTAIM analysis along the two electronic states puts forward that there is no
non-nuclear attractor of the density and no Laplacian basin in the bonding region
corresponding to the transferred electron. Indeed, depending on the
interatomic separation, the electron is located in the lithium or the hydrogen atom but it
is never found in a separate entity.
Therefore, a three-dimensional
representation of these functions cannot be used to
pinpoint the electron position as it jumps between atoms.\newline

\begin{figure}[h]
\begin{center}
\includegraphics[angle=270, scale=0.45]{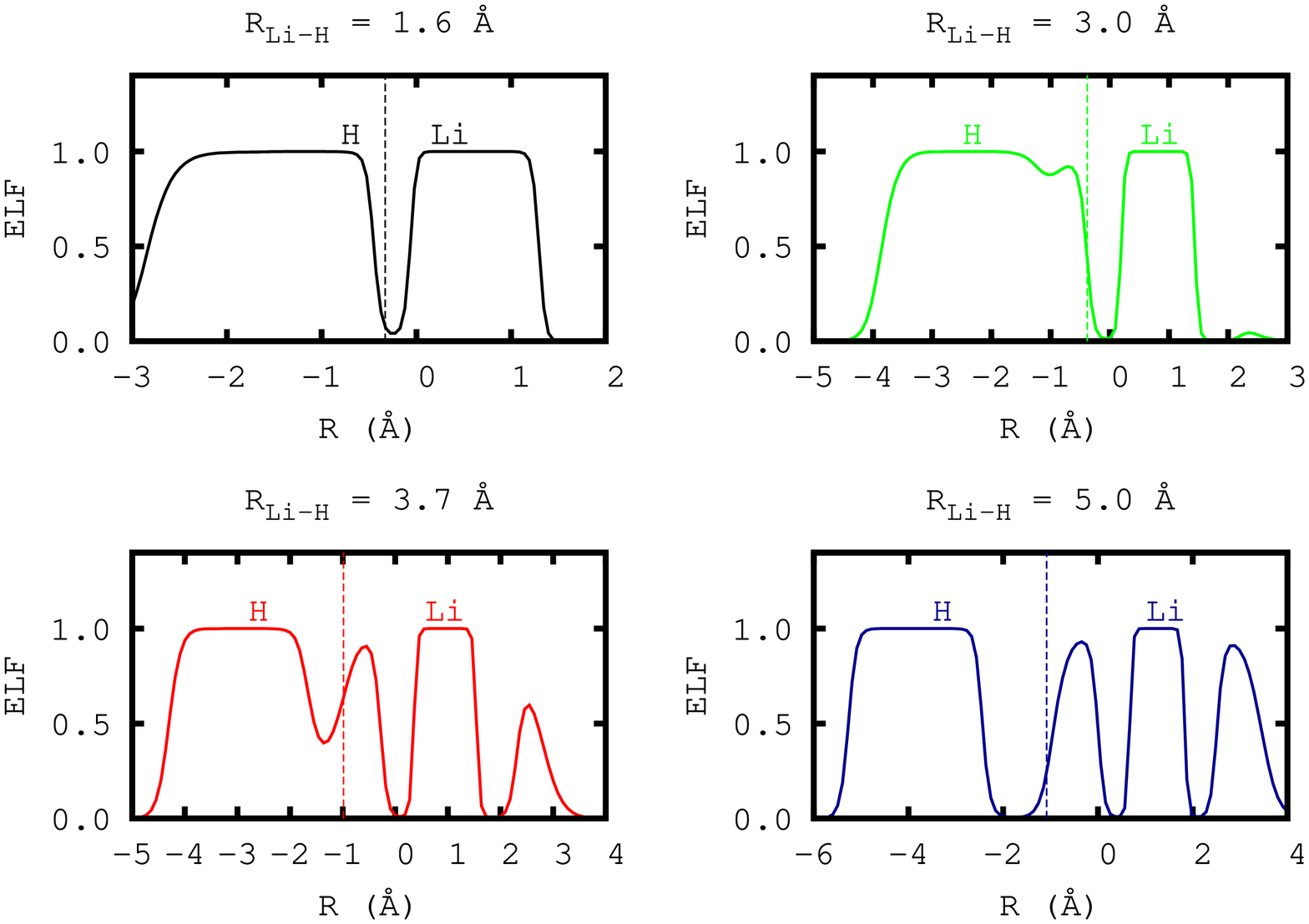}
\end{center}
\caption{The ELF profile of the X$^1\Sigma^+$ state of LiH along the internuclear axis for 
several values of the interatomic distance. The zero is located at the center
of mass and the vertical dashed line indicates the bond critical point. 
The first avoided crossing occurs at about 3\AA.}
\label{fig:lihelf}
\end{figure}

\begin{figure}[h]
\includegraphics[angle=270, scale=0.5]{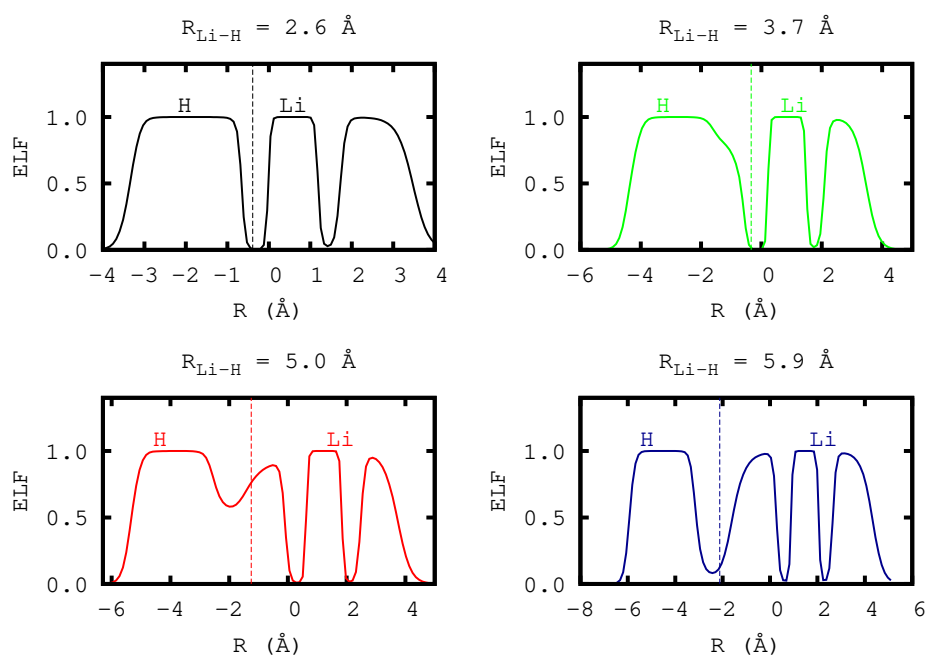}
\caption{The ELF profile of the A$^1\Sigma^+$ state of LiH along the internuclear axis for 
several values of the interatomic distance. The zero is located at the center
of mass and the vertical dashed line indicates the bond critical point.
The first avoided crossing occurs at about 3$\AA$ and the second at 6\AA.}
\label{fig:lihelfes}
\end{figure}

The electron jump process is
difficult to identify by the isosurface representation but it 
is easily monitored by
ELF profiles along the internuclear axis.
Fig.~\ref{fig:lihelf} includes several profiles of the ELF along the molecular axis 
of LiH's X$^1\Sigma^+$ state at
different interatomic distances. A vertical dashed line marks the minimal density 
position between Li and H in the molecular axis. The latter is known in the QTAIM
framework as the bond critical point (BCP), and in these plots
the BCP indicates the atomic boundary
along the internuclear axis. These plots show
that at about the avoided crossing region the valence basin of lithium emerges
at one end of the molecule, whereas the hydrogen valence basin starts to
split into two but it still belongs to the H atomic QTAIM basin.
At 3.3$\AA$ the hydrogen peak splits into two and the
second peak carrying the transferred electron crosses the boundary between
QTAIM atomic basins afterwards.
At 5$\AA$ the process is almost complete and the lithium valence basin is
fully characterized (see the two peaks around the lithium atom
in the one-dimensional plot of Fig.~\ref{fig:lihelf}).
At no point during the dissociation the (one-dimensional) maximum
between Li and H in Fig.~\ref{fig:lihelf} becomes a 
three-dimensional ELF maximum that would give rise to a separate basin.
In this sense, a one-dimensional ELF representation is more useful than 
the usual isosurface picture in order to trace the \textit{electron motion} 
in the reaction mechanism.
Inspection of the A$^1\Sigma^+$ state ELF profile along the molecular axis
reveals the peak formation after the first avoided crossing and the
peak splits and moves towards lithium until it has completely passed the atomic QTAIM
boundary after the second avoided crossing (see Fig.~\ref{fig:lihelfes}).\newline

\begin{figure}[h]
\begin{center}
\includegraphics[angle=270, scale=0.4]{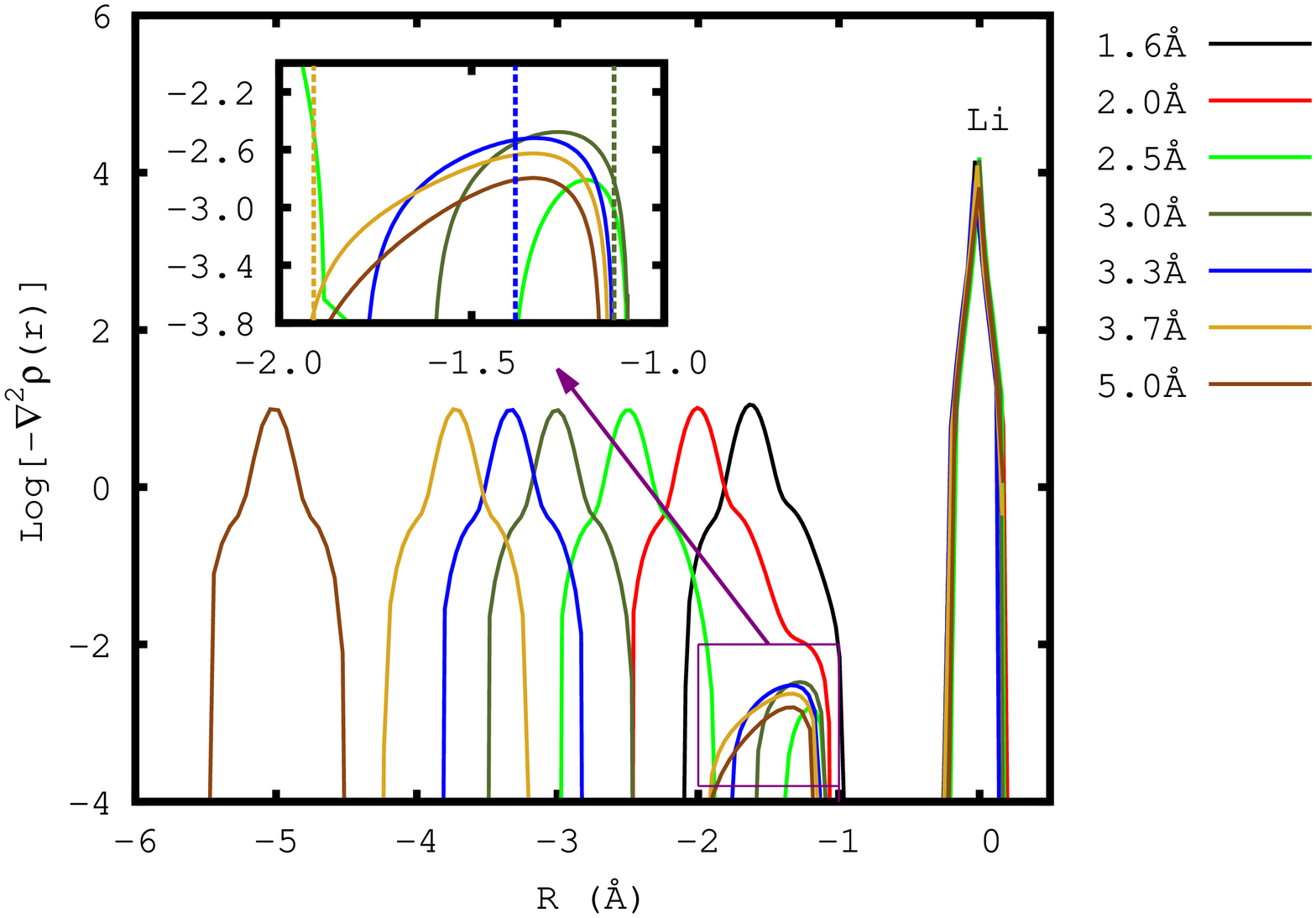}
\includegraphics[angle=270, scale=0.4]{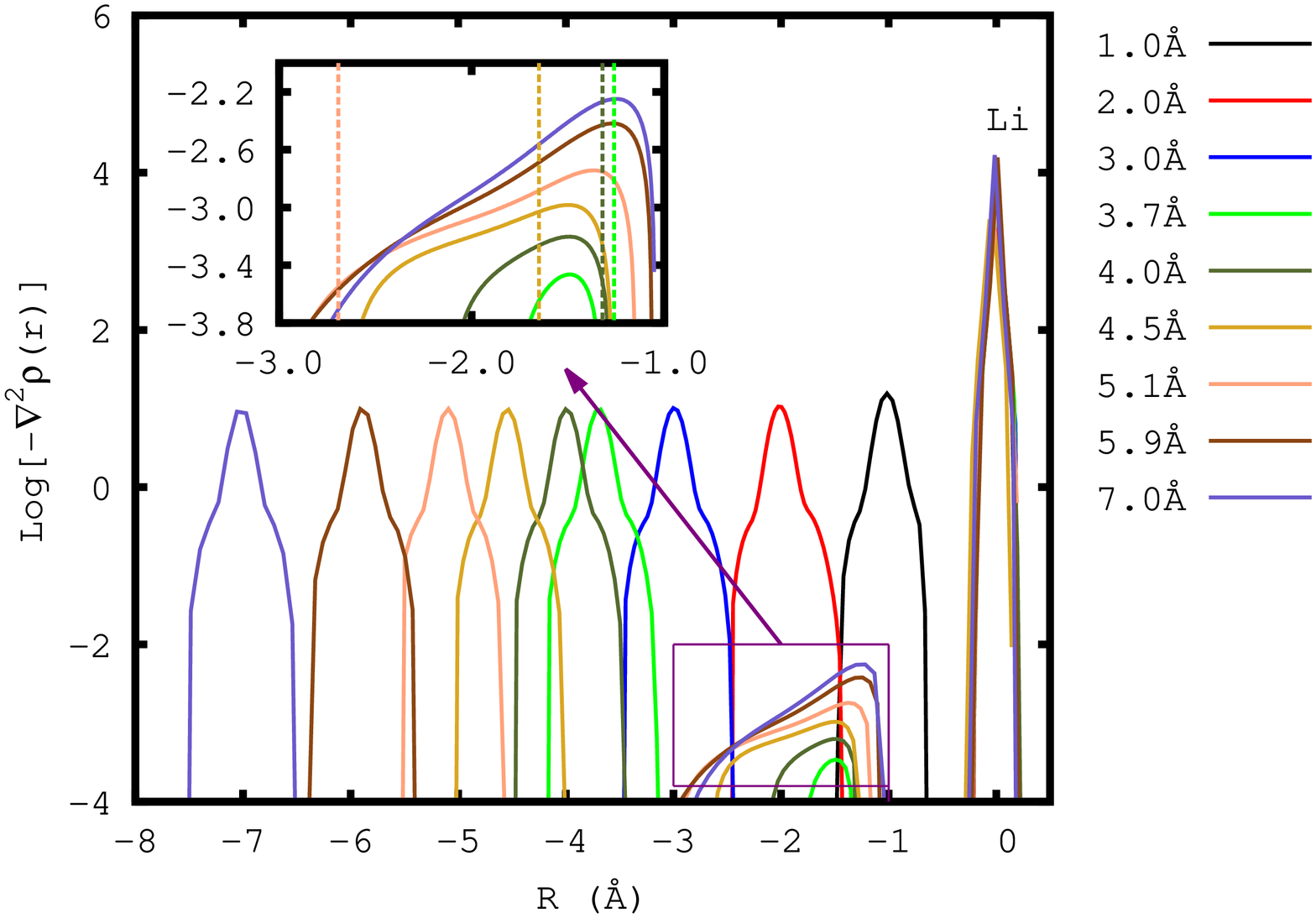}
\end{center}
\caption{The negative values of the Laplacian of the electron density of 
LiH along the internuclear axis 
for several interatomic distances (top X$^1\Sigma^+$, bottom A$^1\Sigma^+$).
All molecules locate Li atom in the z-axis at zero for comparison.
The vertical dashed lines in the inset plot indicate the bond critical points.
Laplacian units are a.u.}
\label{fig:lihlapl}
\end{figure}

Similarly, the molecular-axis scan of the Laplacian of the electron density 
at different interatomic separations reveals the harpooning mechanism.
In Fig.~\ref{fig:lihlapl} we represent the negative
values of the Laplacian of the electron density along the molecular axis 
at different interatomic
distances; we have used logarithmic scale to allow an easy identification
of the peaks and dashed vertical lines to indicate the BCPs.
For the X$^1\Sigma^+$ state at 2$\AA$, the peak corresponding to the hydrogen
atom starts to expand, splitting into two separate peaks after 
2.5\AA.\footnote{Setting the position of the peaks' separation is not
exempt of numerical error. We have set the peaks' separation after 2.5$\AA$ by
neglecting values of the Laplacian below 10$^{-4}$.}
Then, the two peaks separate from each other and, eventually, the new peak 
separates completely from hydrogen peak and becomes the valence basin of lithium.
Interestingly, the new peak containing the transferred electron starts crossing
the QTAIM atomic boundary at the first avoided crossing.
For the A$^1\Sigma^+$ state, the peak between Li and H shows at about $3.7\AA$.
As the distance elongates, the peak broadens and it begins to cross the
QTAIM atomic boundary after 4.0$\AA$ and it is completely transferred when
the molecule passes the second avoided crossing
(see Fig.~\ref{fig:lihlapl} bottom picture). Therefore, according to the
ELF-QTAIM description the electron 
transfer of the A$^1\Sigma^+$ state also takes place
between the two avoided crossings.

\subsection{Other Bond Formations}

In this section we analyze the formation process of other diatomic molecules 
in their ground state from the isolated neutral atoms.
First, we have chosen two additional molecules,
BeH and LiF, which are also formed through an electron harpooning mechanism
due to ionic and covalent diabatic state crossings.
The former consists of Be, which has an ionization potential
larger than Li and, therefore, contributes to a larger energy difference with hydrogen's
electron affinity than LiH. As a result, the electron jump takes places at quite short
distances. Conversely, F has a larger electron affinity than H, giving a small
IP-EA value for the atoms in LiF. The electron jump in LiF thus occurs at large 
interatomic separations but the state-average CASSCF description of the process
underestimates the distance at which the avoided crossing occurs (\textit{vide supra}).
Notwithstanding, it does not affect the qualitative bonding picture of LiF
or the bond mutation process around the avoided crossing.
Second, we analyze H$_2$, O$_2$, N$_2$, F$_2$, He$_2$, BH and CO, which are 
mostly included for comparison. These molecules
present different bonding patterns from weakly bonded to covalent single-,
double-, and triple-bonds, as well as a polar covalent bond. An example of the
latter is the CO molecule and its formation also implies an important electron
reorganization process in 
spite of the fact that the mechanism does not involve electron harpooning.\newline

\begin{figure}[h]
\begin{center}
\includegraphics[angle=270, scale=0.4]{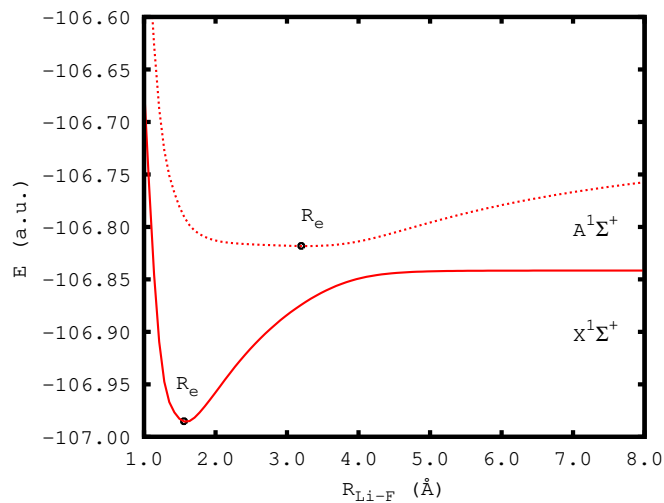}
\end{center}
\caption{Potential Energy Curve (PEC) of the LiF for the ground state (X$^1\Sigma^+$) and
the lowest-lying A$^1\Sigma^+$ excited state. 
Energy in a.u.}
\label{fig:peslif}
\end{figure}

Let us first analyze BeH and LiF. Fig~\ref{fig:peslif} shows the two lowest-lying
$^1\Sigma^+$ states of LiF: the X$^1\Sigma^+$ and A$^1\Sigma^+$ states. 
The dipole moment of both states of LiF
coincides at a distance close to the avoided crossing (ca. $3.6\AA$)
(see Fig~S1 in the Supp. Info). The ESI profile in Fig~\ref{fig:esis} also shows a
maximum around the avoided-crossing interatomic distance for both states. 
The ESI profile
of LiF's A$^1\Sigma^+$ state shows larger (slightly more covalent) values 
than the X$^1\Sigma^+$ state before the avoided crossing
but the rest of the ESI profile is very similar.
The avoided crossing point is not only the geometry at which the ESI is maximum, but
also the point at around which there is a maximal transfer variation and a net transfer
of 0.5 electrons from Li to F as compared to its isolated neutral counterparts.
The ELF and the Laplacian
profiles along the internuclear axis show a separate peak (corresponding to the transferred 
electron from F to Li) that emerges before the avoided crossing and it crosses
the QTAIM atomic boundary around the avoided crossing position (see Figs. S2 and S3).\newline

Unlike LiH and LiF, BeH's X$^2\Sigma^+$ ground state avoided crossing occurs at very 
short distances (ca. 2.2$\AA$~\cite{cooper:84jcp}).
The ESI profile reveals a maximum at this position that is also the
same position at which the transfer variation is maximal and 0.5 electrons have been
transferred from H to Be. 
The ELF and Laplacian
show features similar to those of LiH and LiF, with the peak corresponding to the
electron carried from one atom to another crossing the QTAIM atomic boundary around the
avoided crossing (see Figs S4 and S5).\newline

BH in its X$^1\Sigma +$ state presents a rather large ESI value for an ionic species
($\delta(B,H)=0.6$) and, actually, 
the valence bond description assigns a covalent character to this bond~\cite{miliordos:08jcp}.
The ground state energy curve does not 
present avoided crossings that rapidly could change the bond character from
ionic to covalent or viceversa. Therefore, all evidences suggest a
electron reorganization without harpoon mechanism. 
Notwithstanding, our results show otherwise. The ESI profile is characteristic
of a harpooning mechanism taking place at very short distances (ca. 1.5\AA).
A fact that is confirmed by ELF and Laplacian profiles along the
molecular axis (see Figs S6 and S7), which show that
ELF and Laplacian hydrogen's peak splits into two and passes the atomic
QTAIM boundary from H to B at the vicinity of the ESI maximum,
where the molecule experiences its maximal 
transfer variation and 0.5 electrons have already passed from H to B. Therefore, one is
deemed to conclude that the formation of BH also takes places through a harpoon
mechanism without the existence of an avoided crossing.\newline

According to QTAIM population analysis (see Fig.~\ref{fig:pops}),
carbon withdraws more than one electron from oxygen in the CO molecule.
Unlike the LiH or LiF, the electron-density transfer does not imply a large change in
the dipole moment. Although the shape of the dipole moment profile along 
the dissociation is similar in all electron-transfer reactions, the reaction
in CO (see Fig. S8)
takes places through very small values of the dipole moment (max. 0.3, min. -0.1).
The ESI profile exhibits a smoothly decreasing profile that dies off after 4$\AA$.
There is, therefore, no ESI maximum at stretched distances but the maximal
transfer variation coincides again with the point where 0.5 electrons have been transferred
from C to O.
The Laplacian and the ELF scans along the interatomic axis describe the dissociation
process as the splitting of the bonding peak (corresponding to the triple bond)
into two peaks (see Fig. S9 and S10 in Supp. Info.).\newline

The rest of the molecules are calculated to afford a comparison between 
the above-described electron reorganization processes and other reactions.
The ESI profiles of F$_2$, O$_2$ and N$_2$ exhibit a sigmoidal shape that
Pend\'as and coworkers identify with the breaking of covalent bonds, whereas
the He$_2$ ESI dissociation curve decays exponentially, in line with noncovalent
interactions~\cite{garcia-revilla:11jctc}. The ELF and Laplacian scan
along the molecular axis do not have much interest because they do not
reveal additional features as compared to the isosurface representations
discussed elsewhere~\cite{silvi:05monat}.

\subsection{Information theory indicators}

\begin{figure}[h]
\includegraphics[angle=270, scale=0.25]{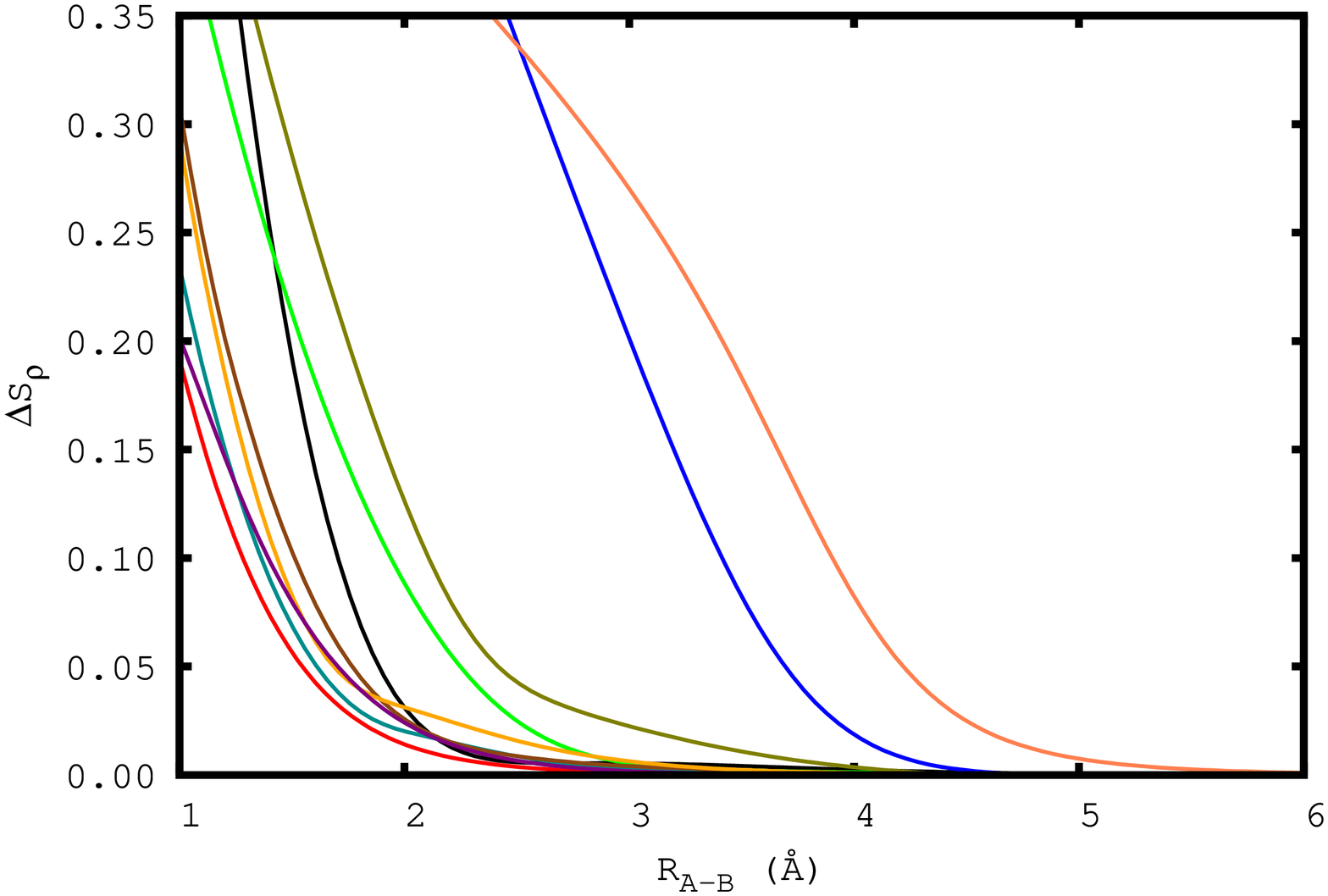}
\includegraphics[angle=270, scale=0.25]{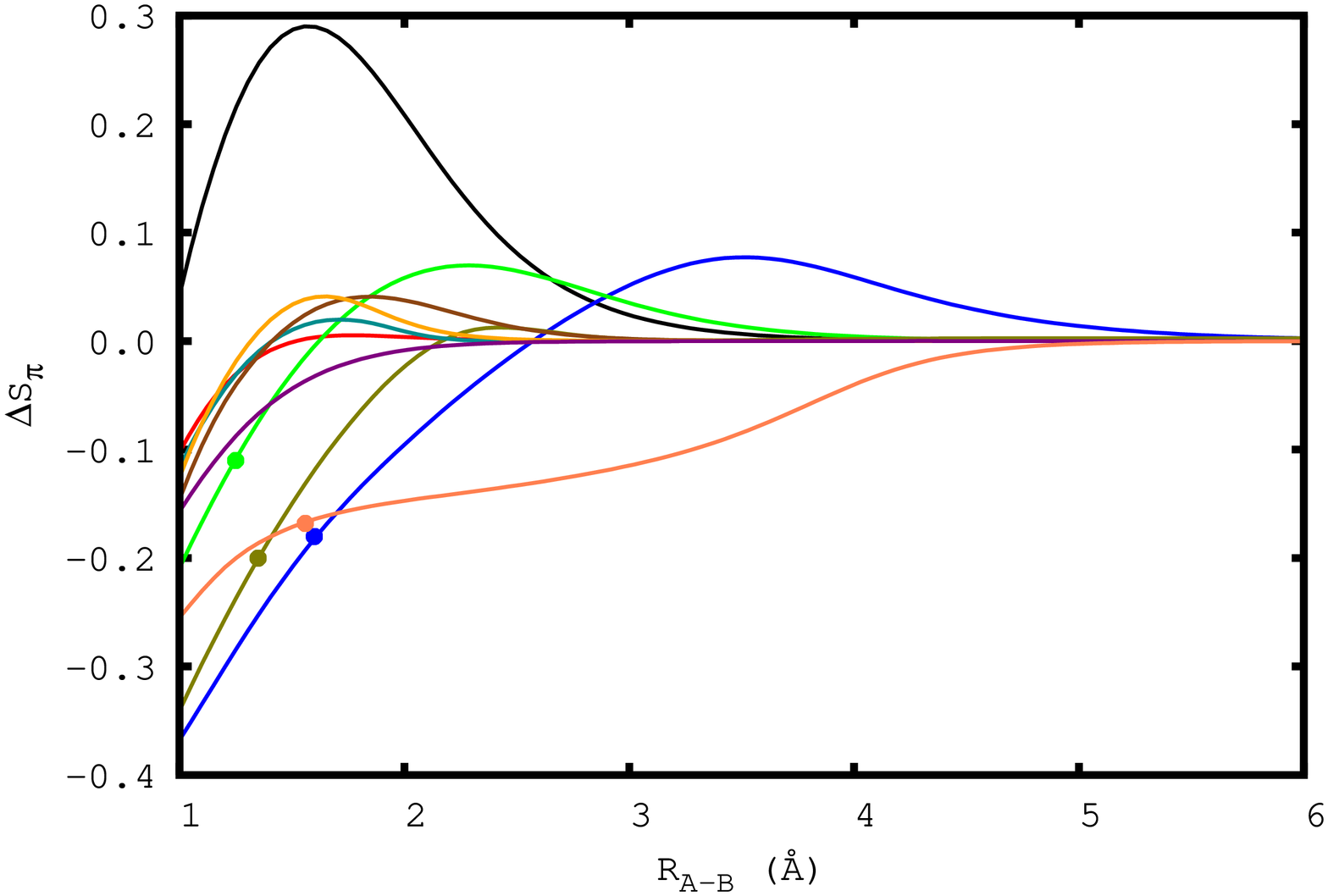}
\includegraphics[angle=270, scale=0.25]{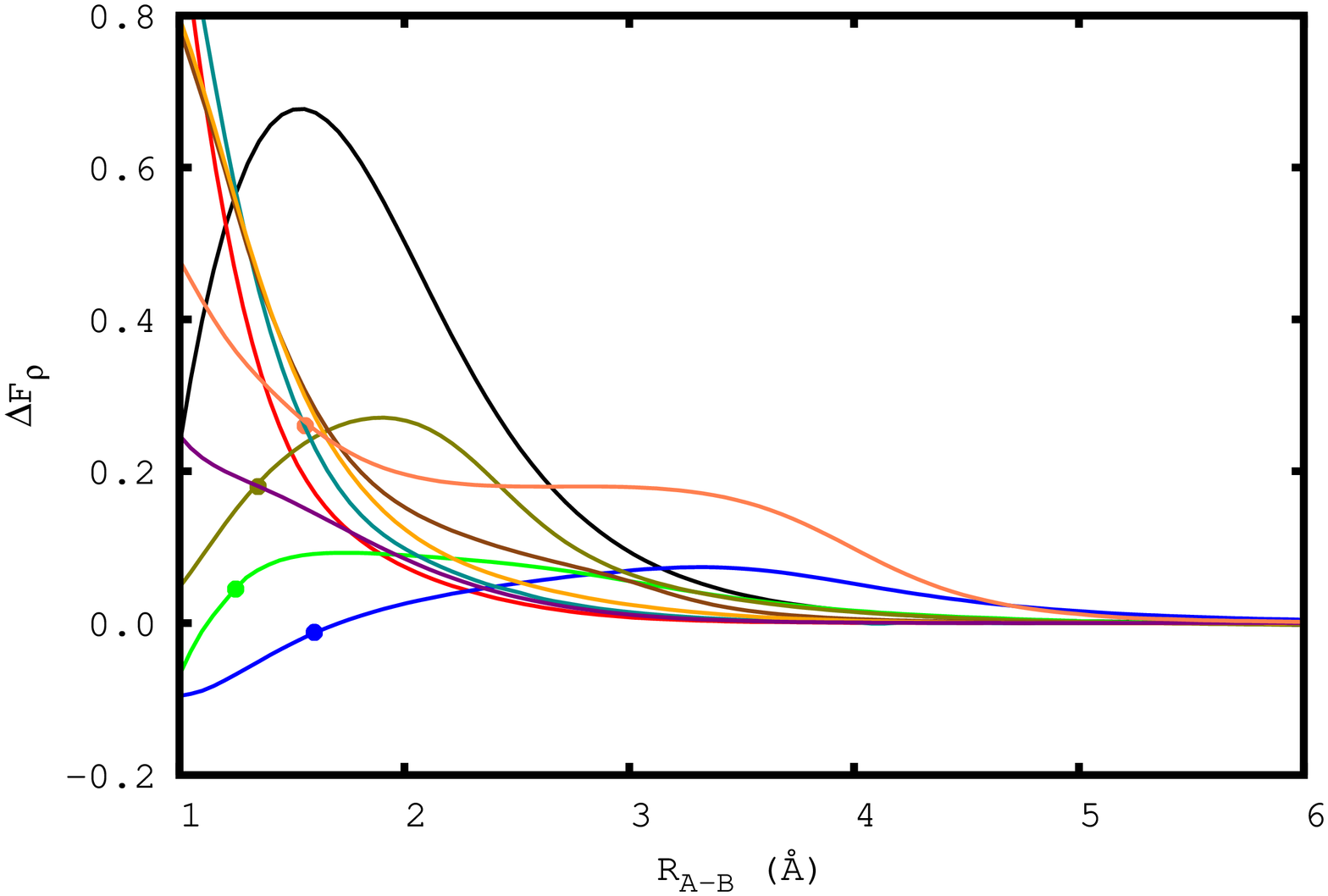}
\includegraphics[angle=270, scale=0.25]{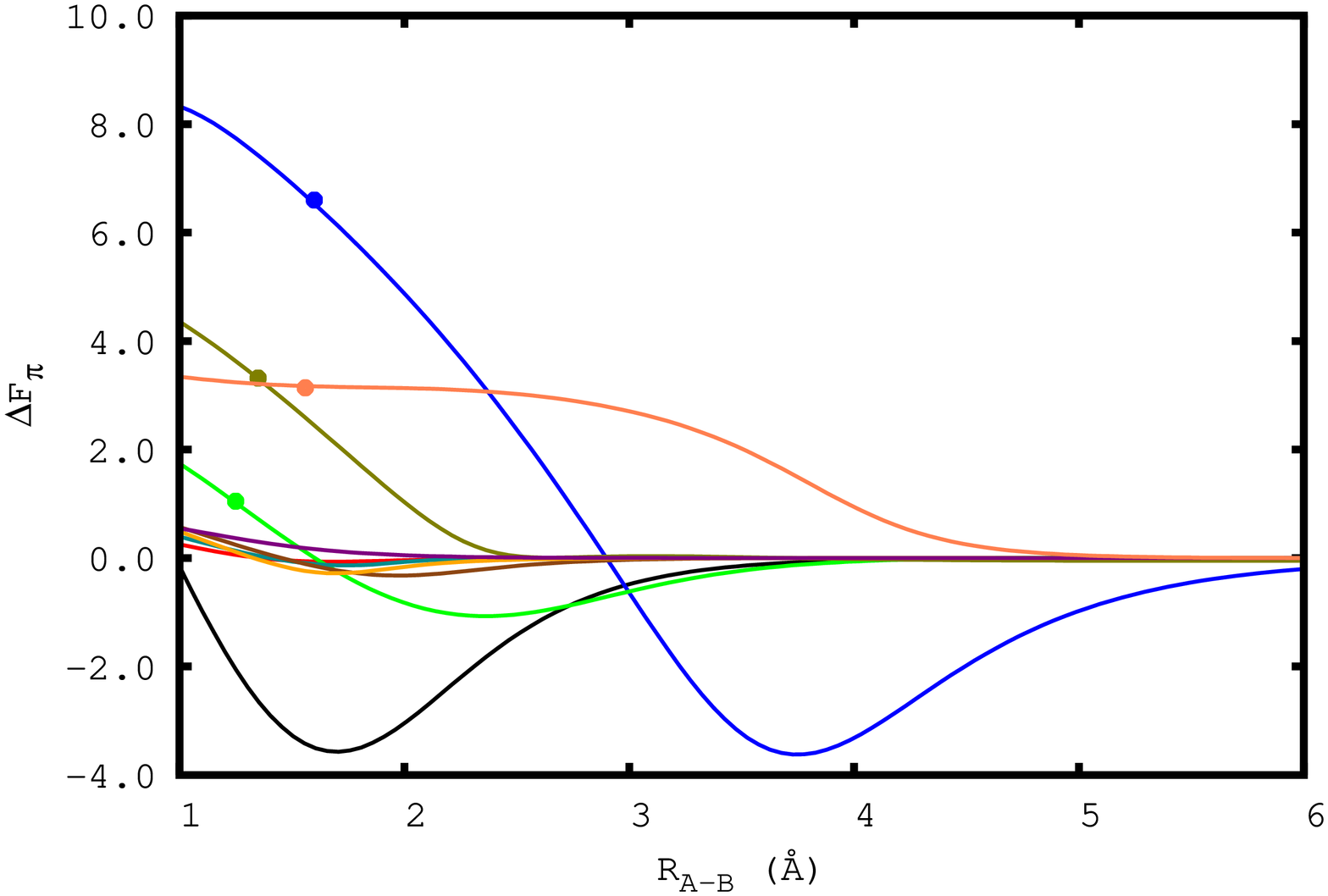}
\begin{center}
\includegraphics[angle=270, scale=0.25]{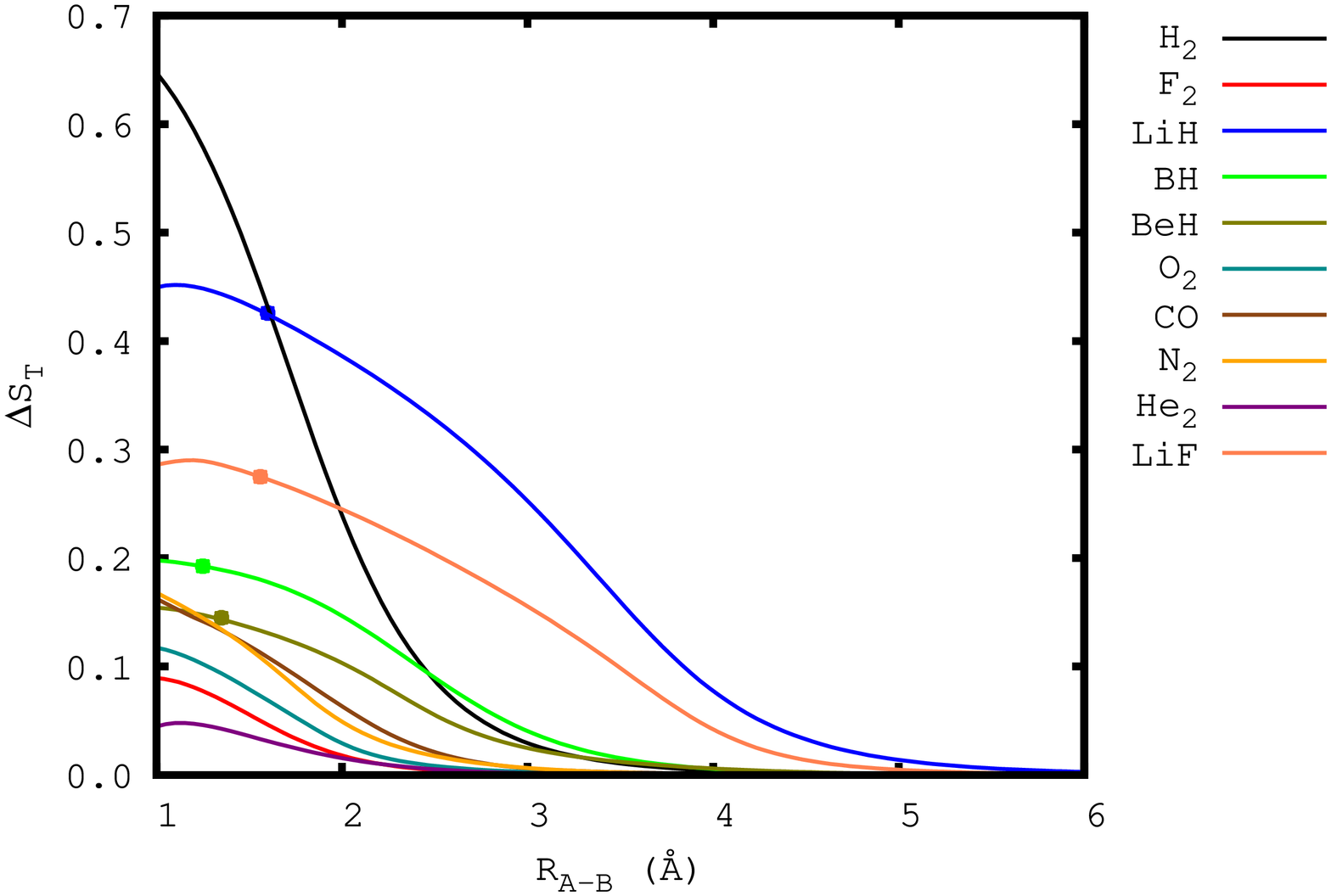}
\end{center}
\caption{From left to right and top to bottom: 
(a) Shannon Entropy in the position space (see Eq.~\ref{eq:sr});
(b) Shannon Entropy in the momentum space (see Eq.~\ref{eq:sp});
(c) Fisher Information in the position space (see Eq.~\ref{eq:fr});
(d) Fisher Information in the momentum space (see Eq.~\ref{eq:fp});
(e) Total Shannon Entropy ($S_T = S_{\rho} + S_{\pi}$); 
plots represent the values at the dissociation limit minus the value at
different interatomic separations
for the series of diatomic molecules taken into study.
Information theory descriptors units are a.u.
}
\label{fig:sr}
\end{figure}

Finally, let us examine the profiles provided by five indicators defined in the
framework of information theory. Fig.~\ref{fig:sr} collects
FI and Shannon indicators (see. Eqns.~\ref{eq:sr}-\ref{eq:fp}) 
at the dissociation limit minus its value at different interatomic separations
($\Delta S_{\rho}, \Delta S_{\pi}, \Delta F_{\rho}, 
\Delta F_{\pi}$ and $\Delta S_T$).\footnote{Please notice 
that these quantities are not size extensive and, therefore, their values can
not be compared with the sum of the isolated fragments.} 
Since the molecules analyzed
undergo different bond formation mechanisms 
one would expect some distinguishing features in the representation
of these indicators along the dissociation curves.\newline

$\Delta S_{\rho}$ shows that 
molecules that undergo an electron-transfer process show larger differences
with respect to its value at infinite separation.
These molecules experience an internal electron reorganization at stretched
distances that is expected to affect its $S_{\rho}$. In this regard,
LiF and LiH experience a long-range electron exchange that explains why
these molecules converge to $\Delta S_{\rho}\rightarrow 0$ at larger
distances than the other molecules. On the other hand, it is not clear
why the H$_2$ profile lies very close to BeH and BH curves or why 
covalent molecules show $\Delta S_{\rho}$ profiles very similar to
He$_2$. Therefore, it does not seem to be an obvious connection between $\Delta S_{\rho}$ 
profiles and the bond formation mechanim in diatomics.
Similar conclusions can be drawn from the inspection of $\Delta S_T$ profiles.\newline

$\Delta F_{\rho}$ shows maximal values 
more or less close to the avoided crossings for LiH, LiF and BeH, whereas
$\Delta F_{\rho}$ decays exponentially for the other molecules except
He$_2$, which shows a somewhat sigmoidal shape. In this sense, the profile
of $\Delta F_{\rho}$ could provide a distinctive shape for each type
of mechanism, excluding H$_2$ whose shape does not match the other
covalent molecules. On the other hand, there does not seem to be a way to
distinguish between single-, double- or triple-bond dissociations.\newline

As far as the magnitudes based on the momentum density ($\Delta S_{\pi}$
and $\Delta F_{\pi}$) go, we can see that 
the molecules involving a electron-transfer harpoon mechanism
present larger
differences with respect to the value at infinite separation.
One more time,
hydrogen molecule is an exception to this rule.

\section{Discussion and Conclusions}

The lowest-lying states of LiH have been widely used to develop
and calibrate many different methods in quantum mechanics. In this paper
we have shown that the electron-transfer processes occurring in these two 
states are a difficult test for chemical bonding descriptors and can be
used with a two-fold purpose: (i) to test new bonding descriptors on
its ability to recognize the harpoon mechanism and (ii) to evaluate the goodness
of new atomic partitions to reproduce the maximal ESI value in several reactions.\newline

Despite the harpoon mechanism involves the electron transfer at large interatomic
distances, this reaction can be mistaken by a regular electron reorganization
process that does not involve a harpooning process. Indeed, in
Fig.~\ref{fig:pops} one finds the population analysis of
BeH, which is formed by the harpoon
mechanism at short interatomic distances. 
The Be population profile could be easily confused
with a regular electron reorganization process as that taking place in CO. 
Therefore, the atomic
population can be used to monitor the electron-density exchange between atoms but it cannot
be always used to discriminate the reaction mechanism. 
In all studied mechanisms, the maximal
transfer variation point along the bond formation occurs when about half electron 
has been transferred from one atom to another. If the process takes places
through a harpoon mechanism, this point of the reaction path coincides with
the avoided crossing (if there is any).
Thus far, the only
bonding descriptor that had actually been shown to identify the harpoon
mechanim from other processes is the ESI~\cite{ponec:05the,matito:07fd}, 
which shows a maximum around the avoided crossing (see Fig.~\ref{fig:esis}).
The existence of a maximum of the ESI does not guarantee the existence of an
avoided crossing as proved by the BH molecule. However, the ESI provides 
a simple tool to identify the existence of
the avoided crossing if there is such point in the potential energy curve.\newline

Some of us have shown in the past that 
atomic populations and multicenter ESI are severely affected by the atomic 
partition~\cite{heyndrickx:11jcc} but, in the big picture, atomic partitions 
do not have a large effect on the relative
two-center ESI values~\cite{matito:05jpca}.\footnote{The only exception are
Hilbert space partitions (such as Mulliken's or L\"owdin's), 
which show spurious results if the calculations
involve a many-body wave function and its exact pair density function~\cite{ramos-cordoba:12pccp}.}
However, Fig~\ref{fig:esis} shows that the presence of a maximum
ESI value in bonds formed by the harpoon mechanism
is a very constrictive test for atomic partitions. The TFVC is among the few
atomic partitions that can pass such test (see Fig. S11), even though 
the ESI values calculated with
this partition fail to identify the bond in LiH's A$^1\Sigma^+$ state as more covalent 
than the bond in the X$^1\Sigma^+$ state at short distances (R=$1-3\AA$).\newline

This work also shows that ELF and Laplacian plotted along the molecular axis at
different interatomic distances permit to monitor the bond formation process better
than its three-dimensional counterparts. Both tools describe the transferred electron
as a one-dimensional peak that moves from one atom to another as the process takes
places. The position of the peak with respect to the QTAIM boundary permits to follow
this motion and shows that the peak passes from one atom to the other at the vicinity
of the avoided crossing, where the ESI value and the transfer variation are maximal.\newline

Global information theory measures based on the electron density seem to provide some distinctive
features for different mechanisms of bond formation
but they do not provide a convincing description that
allows a clear-clut separation between the different processes. Sometimes the descriptors
provide alike profiles for different bonding mechanisms and other times they provide very disparate 
results for similar processes. In addition, 
the results obtained for
the H$_2$ dissociation are not trivial to interpret and suggest that
previous successful results describing chemical reaction paths were, to some extent, fortuitous
~\cite{lopez:09jctc,esquivel:09tca}. The information theory descriptors analyzed
in the atomic regions within the molecule~\cite{geerlings:11pccp,nalewajski:00jpca}
or the orbital entaglement measures~\cite{murg:15jctc} (which can be used to
locate avoided crossings)
could actually provide some insights into the bonding mechanism of molecules but
such study is beyond the scope of this work.\newline

\section*{Acknowledgements}

The authors thanks Prof. Bernard Silvi for providing Ref.~\cite{krokidis:98njc}.

\section*{Funding}
This research has been funded by the MINECO projects CTQ2014-52525-P
and CTQ2014-54306-P,
the Basque Country Consolidated Group Project No. IT588-13,
the FEDER grant UNGI10-4E-801 (European Fund for Regional Development),
the Generalitat de Catalunya (project number 2014SGR931, Xarxa de Refer\`encia en Qu\'imica Te\`orica i Computacional, and ICREA 2014 prize to M.S.) and the predoctoral fellowship
FPU13/00176.
Excellent service by the Centre de Serveis Cient\'ifics i Acad\`emics de Catalunya (CESCA) 
and technical and human support provided by SGI/IZO-SGIker UPV/EHU
are gratefully acknowledged.


\label{lastpage}

\end{document}